\documentclass[aps,12pt,onecolumn]{revtex4-1}
\usepackage{amssymb,amsmath}
\usepackage{hyperref}
\usepackage{color}
\usepackage{graphicx}
\usepackage{ulem}
\usepackage{mathbbol}
\usepackage{epsfig}

\newcommand{\sinc}{\mathop{\mathrm{sinc}}\nolimits}
\def\beq{\begin{equation}}
\def\eeq{\end{equation}}

\begin{document}ñ
\title{Understanding and enhancing superconductivity \\ in FeSe/STO by quantum size effects}
\author{Bruno Murta}
\author{Antonio M. Garc\'ia-Garc\'ia}
\email{amg73@cam.ac.uk}
\affiliation{Cavendish Laboratory, University of Cambridge, JJ Thomson Av., Cambridge, CB3 0HE, UK}
\date{\today}
\begin{abstract}
	Superconductivity in one-atom-layer iron selenide (FeSe) on a strontium titanate (STO) substrate is enhanced by almost an order of magnitude with respect to bulk FeSe. There is recent experimental evidence suggesting that this enhancement persists in FeSe/STO nano-islands. More specifically, for sizes $L \sim 10$ nm, the superconducting gap is a highly non-monotonic function of $L$ with peaks well above the bulk gap value. This is the expected behavior only for weakly-coupled metallic superconductors such as Al or Sn. Here we develop a theoretical formalism to describe these experiments based on three ingredients: Eliashberg theory of superconductivity in the weak coupling limit, pairing dominated by forward scattering and periodic orbit theory to model spectral fluctuations. We obtain an explicit analytical expression for the size dependence of the gap that describes quantitatively the experimental results with no free parameters. This is a strong suggestion that superconductivity in FeSe/STO is mediated by STO phonons. We propose that, since FeSe/STO is still a weakly coupled superconductor, quantum size effects can be used to further enhance the bulk critical temperature in this interface.

\end{abstract}

\maketitle

\section{Introduction}

Bulk iron selenide (FeSe) has a relatively low critical temperature $T_c \sim 8$ K with respect to other iron-based superconductors. Surprisingly, a much higher critical temperature $T_c > 40$ K was reported \cite{qing2012,zhang2014a} in a single atomic layer of FeSe (with a capping layer) on a strontium titanate (STO) substrate.
Additional scanning tunneling microscopy (STM) measurements \cite{zhang2014}, in-situ transport results \cite{ge2015} using a four-probe STM technique, and ARPES \cite{lee2014} experiments have not only confirmed this enhancement but also pointed to an even higher critical temperature ${\rm T_c} \sim 100$ K in the absence of a capping layer.  

Interestingly, in multi-layer FeSe heterostructures \cite{wang2015} ${\rm T_c}$ decreases sharply as the number of FeSe layers increases. However, the energy gap, as measured by STM techniques, is non-zero only for a single FeSe layer. This is a clear indication that the substrate plays a key role in the enhancement of ${\rm T_c}$. Indeed, it is well established by now that the Fermi surface of bulk FeSe and the one in FeSe/STO are qualitatively different: only the latter is particle doped. Charge transfer from the substrate to the FeSe layer is expected to enhance superconductivity as it increases the number of carriers available. Nevertheless, this additional charge is not enough to justify such a dramatic enhancement of ${\rm T_c}$ \cite{lee2014,coh2015,gor2015}.


A recent ARPES experiment \cite{lee2014} has revealed the existence of strongly peaked replica bands approximately $100$ meV away from the original electron-like and hole-like bands in FeSe/STO. Given that STO has a very flat optical phonon band precisely centered around $100$ meV \cite{li2014,Choudhury2008}, and since these oxygen vibrational modes are widely separated from other phonon modes, the  occurrence of these replica bands is likely due to the coupling between 3d FeSe electrons and the optical oxygen phonon branch in the STO substrate. This novel forward scattering mechanism \cite{rademaker2015,lee2015}, which had previously been found to be relevant in other superconductors \cite{kulic1994,perali1996,santi1996,perali1998}, has in principle the potential to explain the high critical temperature observed in experiments. 


Indeed, although different theoretical  models \cite{gor2015,rademaker2015} have already been employed to model FeSe/STO, the approach of \cite{rademaker2015} is perhaps the most promising one as values close to the experimental critical temperature were obtained by considering forward scattering \cite{kulic1994,perali1996,santi1996,perali1998,lee2015} as the sole superconductivity mechanism. Unlike the usual BCS prediction, the critical temperature is approximately proportional to both the Debye energy and the electron-phonon coupling constant. We note that this approach employs the conventional Eliashberg formalism that assumes that Migdal's theorem holds. For this to happen the Fermi energy must be larger than the Debye energy. 
In FeSe/STO the Debye energy is of the order of the Fermi energy, but corrections to the Eliashberg formalism \cite{grimaldi1995} due to deviations from Migdal's theorem are still small in the limit of weak coupling $\lambda \leq 0.25$ that seems to describe the FeSe/STO experimental results. Indeed, the results of a recent calculation \cite{wang2016} of vertex corrections in FeSe/STO provide further support to the applicability of the Eliashberg formalism.

Recent STM measurements \cite{wang2015} in one-layer FeSe/STO nano-islands of typical size $L \sim 10$nm have shown that the superconducting gap is a highly non-monotonic function of the grain size. Even small changes in the grain size induce large variations of the gap with peaks and valleys that deviate substantially ($\sim 40-50\%$) from the bulk limit. 
This is hardly an exception as there are already a plethora of theoretical and experimental studies \cite{blatt1963,bose2005,brun2009,carbillet2016,cohen1968,deutscher1973,giaver1968,guo2004,perali1996a,pracht2016,zgirski2005,strongin1970,rao1984} that have shown the importance of size effects in superconductivity when one or more dimensions is reduced to the nano-scale (see \cite{bose2014} for an excellent review focused on superconductivity nano-grains). Of special importance in our analysis is the experimental observation of strikingly similar  effects \cite{bose2010} in nano-grains of conventional metallic superconductors such as Al and Sn. Its origin is well understood \cite{parmenter1968,gladilin2002,heiselberg2003,bose2005,shanenko2006,kresin2006,croitoru2007,garcia2008,garcia2011,brihuega2011,araujo2011}: fluctuations of the spectral density around the Fermi energy, enhanced by spectral degeneracies (shell effects), make the gap sensitive to the grain size. Bardeen-Schieffer-Cooper (BCS) theory is enough to model quantitatively these quantum-size deviations from the bulk limit. In standard BCS theory these effects are especially pronounced for sizes much smaller than the superconducting coherence length of the material.
However, its observation in FeSe/STO comes as a total surprise. The coherence length in FeSe/STO is of the order of the grain size and forward scattering suppresses quantum size effects as it restricts the phase space available for pairing. The only possible explanation is that superconductivity in FeSe/STO is not BCS-like, namely the gap or $T_c$ do not depend exponentially on the electron-phonon coupling constant, and that deviations from perfect forward scattering are sufficiently strong.  

Here we propose a theoretical model that describes quantitatively these quantum size effects, thus shedding light on the bulk FeSe/STO superconductivity mechanism. More specifically, we combine semiclassical techniques with the Eliashberg theory of superconductivity in the weak-coupling limit in order to describe theoretically quantum size effects in superconductors with strong forward scattering. We then show that our model describes quantitatively size effects in FeSe/STO nano-islands without the need of any fitting parameter. This is a strong indication that high  $T_c$ superconductivity in FeSe/STO is mostly caused by pairing of FeSe electrons mediated by STO phonons. Finally, we also argue that, as in granular metallic superconductors \cite{mayoh2014} and thin films \cite{blatt1963,perali1996a},  further enhancement of superconductivity is possible by nano-engineering of FeSe/STO nano-grains to form a bulk material.

\section{Results}
 We study quantum size effects in FeSe/STO by combining Eliashberg theory \cite{carbotte2008,abrikosov1975} and forward scattering \cite{kulic1994,perali1996,santi1996,perali1998} with a semiclassical analysis of size effects \cite{garcia2008} based on periodic orbit theory.
 In the bulk limit this problem has already been investigated in detail \cite{lee2015,rademaker2015, wang2016} where it was proposed that forward scattering  could be the main mechanism for the enhancement of superconductivity. Here we study specifically how forward scattering modifies quantum size effects in FeSe/STO. 
 



Within the Eliashberg theory \cite{abrikosov1975,carbotte2008} of superconductivity, the electron self-energy due to the electron-phonon interaction in the weak-coupling limit \cite{rademaker2015} is given by:

\begin{equation}
	\Delta(\textbf{k}, i\omega_n) = \frac{-1}{N \beta}\sum_{\textbf{q}, m} |g(\textbf{k}, \textbf{q})|^2D^{(0)}(\textbf{q}, i\omega_n - i\omega_m) \frac{\Delta(\textbf{q+k}, i\omega_m)}{\omega_m^2 + \epsilon_{\textbf{k} + \textbf{q}}^2 + \Delta^2(\textbf{q+k}, i\omega_m)}
\label{eq:Eliashberg_gap_0}
\end{equation}

\noindent where $\Delta(\textbf{k}, i\omega_n)$ is the gap function, D$^{(0)} (\textbf{q}, i\omega_m) = -2\omega_D / ( \omega_D^2 + \omega_m^2)$ is the bare phonon propagator (assuming a flat phonon mode of Debye energy $\omega_D, \hbar = 1$) and $|g(\textbf{k}, \textbf{q})|$ is the matrix element that describes the electron-phonon interaction. $\epsilon_k$ is the dispersion of the electron (relative to the chemical potential $\mu$), N is the number of momentum grid points, $\beta = 1/k_B T$ is the inverse temperature and $\omega_n = (2n + 1)\pi/\beta$ is a Matsubara frequency.

The assumption that the superconducting properties of FeSe/STO can be described by considering only the phonon-mediated pairing channel in the weak-coupling limit $\lambda \leq 0.3$ \cite{rademaker2015} requires forward scattering \cite{kulic1994,perali1996,santi1996,perali1998} to be included in the model. Replacing $\lambda = 0.3$ and $\omega_D = 100$ meV in the usual BCS expression $\Delta_0 = 2 \omega_D \exp(-1/\lambda)$ gives a bulk gap of only 7 meV, which is far from the experimentally measured 16.5 meV \cite{li2015}. However, solving the Eliashberg momentum-dependent equations for low-momentum transfer gives a gap linear in both the Debye energy and the coupling constant, which would allow the bulk gap to be obtained for a Debye energy of the expected order of magnitude for a small $\lambda$.

 Under the assumption of strong forward scattering, only electrons close to the Fermi level are involved in the pairing. 
 Therefore we assume that pairing occur only at the Fermi level. Another argument in favor of this approximation is that we aim to model experiments \cite{li2015} where theoretical results are compared to the average of the experimental value of the spectroscopic gap measured in different positions of the grain which is closely related to fixing the momentum $k$ to be the Fermi momentum. 
 Other calculations \cite{mayoh2014,shanenko2006} in conventional superconducting nano-grains have shown that the magnitude of mesoscopic effects is not substantially altered by including the $k$ dependence provided that the effective number of states subjected to pairing is not substantially altered. 
 Based on similar arguments we also neglect any angular dependence of $k$ at $k_F$. 
We note that recent theoretical \cite{wang2016}  and experimental results \cite{zhang2016} suggest that, in contrast with previous claims in the literature, the angular dependence must be taken into account for a quantitative description of the gap in FeSe/STO. However, we believe that by averaging over $k$ we would get qualitatively similar results for the mesoscopic fluctuations we are interested in. A more detailed analysis would obscure our main goal, which is making an analytical and parameter-free estimation of the strength of mesoscopic fluctuations in this material. In summary, we assume $|\textbf{k}| \approx k_F$ in (\ref{eq:Eliashberg_gap_0}):
 

\begin{equation}
	\Delta(i\omega_n) = \frac{-1}{N \beta}\sum_{\textbf{q}, m} |g(\textbf{q})|^2D^{(0)}(\textbf{q}, i\omega_n - i\omega_m) \frac{\Delta(i\omega_m)}{\omega_m^2 + \epsilon_{\textbf{k}_F + \textbf{q}}^2 + \Delta^2(i\omega_m)}
\label{eq:Eliashberg_gap}
\end{equation}

The extreme case of low-momentum transfer corresponds to perfect forward scattering, for which no momentum transfer is allowed and hence the matrix element can be written as a Kronecker Delta function. In this limit the bulk gap is found to be $\Delta_0 \approx \frac{2\lambda}{2 + 3 \lambda} \ \omega_D$ \cite{wang2016}. As expected, the expression for the bulk gap is linear in $\lambda$ and $\omega_D$. For $\lambda = 0.22$ and $\omega_D = 100$ meV we get a bulk gap of $\sim$ 16 meV.

However, it is clear that within this perfect forward scattering limit no corrections due to quantum size effects can be expected. Indeed, the fluctuations arising from the quantisation of the energy levels are due to the variation of the number of states that contribute to the interaction as the area of the grain is changed; such change cannot be observed in this case because the Kronecker delta picks a single momentum state for the interaction. As a result, we must consider a finite cut-off in order to observe fluctuations.

In order to mimic the experimental situation, we must therefore consider the case of forward scattering with a finite width \cite{perali1996}. The matrix element may be written as $|g(\textbf{q})|^2 = N g_0^2 h(\textbf{q}) = N \lambda \omega_D^2 h(\textbf{q})$, where $h(\textbf{q})$ gives the functional form of the cut-off. For example, Rademaker et al. \cite{rademaker2015} considered an exponentially decaying cut-off $h(\textbf{q}) = e^{-|\textbf{q}| / q_0}$. Keeping a general form of the cut-off function, equation (\ref{eq:Eliashberg_gap}) becomes:

\begin{equation}
\Delta(i\omega_n) = \frac{2 \omega_D^3 \lambda}{\beta}\sum_{\textbf{q}, m} \frac{h(\textbf{q})}{\omega_D^2 + (\omega_n - \omega_m)^2} \frac{\Delta(i\omega_m)}{\omega_m^2 + \epsilon_{\textbf{k}_F + \textbf{q}}^2 + \Delta^2(i\omega_m)}
\label{eq:Eliashberg_gap_numerical}
\end{equation}

Using the ansatz $\Delta(i\omega_n) = \Delta / (1 + (\omega_n / \omega_D)^2)$ \cite{rademaker2015} and setting n = 0 so that $\omega_n \ll \omega_D$ and therefore $\omega_D^2 + (\omega_n - \omega_m)^2 \approx \omega_D^2 + \omega_m^2$, equation (\ref{eq:Eliashberg_gap_numerical}) becomes:

\begin{equation}
1 =  \frac{2\lambda \omega_D^5}{\beta} \sum_{\textbf{q},m}  \frac{h(\textbf{q})}{(\omega_m^2 + \epsilon_{\textbf{k}_F + \textbf{q}}^2) [\omega_D^2 + \omega_m^2]^2 + \omega_D^4 \ \Delta^2}
\label{eq:Eliashberg_gap_numerical_2}
\end{equation}

The Matsubara frequency summation in equation (\ref{eq:Eliashberg_gap_numerical_2})

\begin{equation}
\frac{1}{\beta} \sum_{m}  \frac{1}{(\omega_m^2 + \epsilon_{\textbf{k}_F + \textbf{q}}^2) [\omega_D^2 + \omega_m^2]^2 + \omega_D^4 \ \Delta^2}
\end{equation}

\noindent can be solved by contour integration before considering the sum over momentum. Assuming $\epsilon_{\textbf{k}_F + \textbf{q}} \ll \Delta_0 \ll \omega_D$ for the range of $\textbf{q}$ considered \footnote{As noted in the description of the terms involved in equation (1), the electron dispersion $\epsilon_k$ is measured relative to the Fermi level, so $\epsilon_{k_F} = 0$. Hence, for $\epsilon_{\textbf{k}_F + \textbf{q}} \ll \Delta_0$, we require $|\textbf{q}|$ to be small, which is indeed the case if we impose a sharp cut-off in $\textbf{q}$.}, the approximate poles of the integrand are $\omega_m = \pm i \sqrt{\Delta^2 + \epsilon^2}, \pm i(\omega_D - \Delta/2), \pm i(\omega_D + \Delta/2)$. After summing over Matsubara frequencies, equation (\ref{eq:Eliashberg_gap_numerical}) becomes:

\begin{equation}
1 = \lambda \omega_D \sum_{\textbf{q}} h(\textbf{q}) \Big( \frac{1}{\sqrt{\epsilon_{\textbf{k}_F + \textbf{q}}^2 + \Delta^2}} - \frac{3}{2\omega_D}\Big).
\label{eq:numerical}
\end{equation}

For an arbitrary cut-off $h(\textbf{q})$, equation (\ref{eq:numerical}) can only be solved numerically. However, in order to study the deviations from the perfect forward scattering limit analytically, and for the sake of simplicity as well, we assume a sharp cut-off
so that $h(\textbf{q}) = 0$ everywhere except for $\textbf{q}$'s within the interval $ \epsilon_F -\epsilon_0 <\epsilon_{\textbf{k}_F + \textbf{q}} < \epsilon_F + \epsilon_0$ where $h(\textbf{q}) = \frac{2\pi}{Aq_0^2}$ with $\epsilon_0 = \hbar^2 q_0^2/2m^*$, $m^*$ is the effective electron mass, $q_0 \sim C/a$, $a$ is the lattice constant of FeSe and C $\sim \mathcal{O}(1)$.  The chosen value of the cutoff $h(\textbf{q}) = \frac{2\pi}{Aq_0^2}$ ensures that the perfect forward scattering limit is recovered for $q_0 \to 0$. 
%

By converting the sum over momentum states into an integral over energy about the chemical potential, (\ref{eq:numerical}) may be rewritten as:

\begin{equation}
	\begin{split}
	1 & = \lambda \omega_D \frac{2 \pi}{A q_0^2} \sum_{|\textbf{q}| < q_0} \Big( \frac{1}{\sqrt{\epsilon_{\textbf{k}_F + \textbf{q}}^2 + \Delta^2}} - \frac{3}{2\omega_D}\Big) \\ 
	& = \frac{\lambda \omega_D}{2 \epsilon_0 \nu_{TF}(0)} \int_{\epsilon_F -\epsilon_0}^{\epsilon_F+\epsilon_0} d\epsilon \ \nu(\epsilon) \Big( \frac{1}{\sqrt{(\epsilon -\epsilon_F)^2 + \Delta^2}} - \frac{3}{2\omega_D}\Big)
	\end{split}
\label{eq:Eliashberg_gap_energy}
\end{equation}

\noindent where $\nu(\epsilon)$ is the density of states at energy $\epsilon$ and $\nu_{TF}(0)=\frac{Ak_F^2}{4\pi\epsilon_F}$ is the bulk density of states at the Fermi energy.

We note that the overlap integrals between the single-particle wavefunctions, which arise from the matrix element, were ignored, since their contribution to the finite size fluctuations of the gap is small \cite{garcia2011}. Therefore, the only correction due to quantum size effects that we consider is the quantisation of the energy levels through the semiclassical expansion \cite{Brack,Gutzwiller} of the spectral density,  
\begin{equation}
\nu(\epsilon) = \nu_{T_F}(0) (1 + \overline{g}(0) + \tilde{g}(\epsilon)).
\end{equation}

\noindent where

\begin{equation}
	 \overline{g}(0) = \pm \frac{\mathcal{L}}{2k_FL^2}
\end{equation}

\begin{equation}
	\begin{split}
	\tilde{g}(\epsilon) & = \ \tilde{g}_{1,2}^{(2)}(\epsilon) - \frac{1}{2} \sum_{i}\tilde{g}_{i}^{(1)}(\epsilon) \ = \\
	& = \ \sum_{L_n \neq 0}^{\infty} J_0(k(\epsilon)L_{n}^{1,2}) -  \sum_{i=x, y} \frac{2L_i}{k_F L^2}\sum_{L_n \neq 0}^{\infty}\cos(k(\epsilon)L_n^{i})
	\end{split}
\end{equation}

\noindent where the plus and minus signs in $\overline{g}(0)$, the Weyl term,  correspond to Neumann and Dirichlet boundary conditions, respectively, $L_x = \alpha L$ and $L_y = L / \alpha$ are the sides of the rectangle ($\alpha > 1$), $\ L^2 = L_x L_y$ is the area, $\mathcal{L} = 2( L_x + L_y)$ is the perimeter and $k_F = \sqrt{2m^{*} \epsilon_F}/\hbar$ is the Fermi wavevector. $J_0$ is the zeroth-order Bessel function of first kind, $\ L_n^{1,2} = 2 \sqrt{L_x^2n^2 + L_y^2m^2}$ is the length of the periodic orbit (n,m) and $L_n^{i} = 2nL_i$ is the length of a single-integer periodic orbit. $\tilde{g}_{1,2}^{(2)}(\epsilon)$ is of $ \mathcal{O}(1/\sqrt{k_FL})$, whereas $\overline{g}(0)$ and $\tilde{g}_{i}^{(1)}$ are both of $ \mathcal{O}(1/k_FL)$. 

Replacing the spectral density by the expression above in (\ref{eq:Eliashberg_gap_energy}) and expanding the gap as

\begin{equation}
\Delta(L) = \Delta_0(1 + f_{1/2} + f_{1}) 
\label{eq:deltaL}
\end{equation} 

\noindent where $f_{i}$ stand for corrections to the gap of order $(k_FL)^{-i}$, the gap equation is solved order by order in $(k_FL)^{-i}$. A detailed derivation of the finite size corrections is presented in Appendix A. Here we only present the highlights of the calculation and state the final results.  
 
The zeroth-order term in $(k_FL)^{-i}$ equality gives the bulk gap for a finite width $\epsilon_0$ of the phonon spectrum:
\begin{equation}
1 = \frac{\lambda \omega_D}{2 \epsilon_0} \int_{-\epsilon_0}^{\epsilon_0} d\epsilon \Big( \frac{1}{\sqrt{\epsilon^2 + \Delta_0^2}} - \frac{3}{2\omega_D}\Big)
\label{eq:bulk_gap_2}
\end{equation}

This integral is evaluated exactly to give:

\begin{equation}
\Delta_0 = \frac{\epsilon_0}{\sinh \left((1/\lambda+3/2)\frac{\epsilon_0}{\omega_D}\right)} \label{eq:bulkgap}
\end{equation}

As expected, the zeroth-order term (i.e. the bulk gap) in this expansion in the small parameter 1/$k_FL$ coincides with the result in the perfect ($\epsilon_0 \to 0$) forward scattering limit \cite{wang2016}. Since for FeSe/STO $\epsilon_0 \ll \Delta_0 \ll \omega_D$ corrections to this limit are expected to be rather small. 

A straightforward calculation (see appendix A for details) results in the following expression for the leading finite size correction:

\begin{equation}
	f_{1/2} = 
	\frac{\int_{-\epsilon_0}^{\epsilon_0} d\epsilon \ \tilde{g}_{1,2}^{(2)}(\epsilon) \ \Big( \frac{1}{\sqrt{\epsilon^2 + \Delta_0^2}} - \frac{3}{2\omega_D}\Big)}
	{\Delta_0^2 \int_{-\epsilon_0}^{\epsilon_0} d\epsilon \frac{1}{(\Delta_0^2 + \epsilon^2)^{3/2}}}
	\label{eq:correction_2}
\end{equation}

Considering the numerator first, using the asymptotic limit of the Bessel function $J_0(x) = \sqrt{2/ \pi x} \ \cos(x - \pi/4)$, expanding the wavevector $k(\epsilon) = k_F (1 + \epsilon / (2\epsilon_F))$ (where $\epsilon_F = \hbar^2 k_F^2 / 2m$ is the Fermi energy) and solving the energy integral within the limit $\epsilon_0 \ll \Delta_0$, the numerator in (\ref{eq:correction_2}) becomes:

\begin{equation}
	\int_{-\epsilon_0}^{\epsilon_0} d\epsilon \ \tilde{g}_{1,2}^{(2)}(\epsilon) \ \Big( \frac{1}{\sqrt{\epsilon^2 + \Delta_0^2}} - \frac{3}{2\omega_D}\Big) = 
	2 \epsilon_0 \Big(\frac{1}{\Delta_0} - \frac{3}{2\omega_D}\Big) \sum_{L_n \neq 0}^{\infty} J_0(k_F L_n) \sinc(L_n / \xi)
	\label{eq:numerator}
\end{equation}

\noindent where $\sinc(x) \equiv \sin(x) / x$ and $\xi = \frac{2\epsilon_F}{k_F \epsilon_0}$ plays the role of coherence length. Therefore contributions from periodic orbits much greater than $\xi$ are strongly suppressed.

\begin{figure}
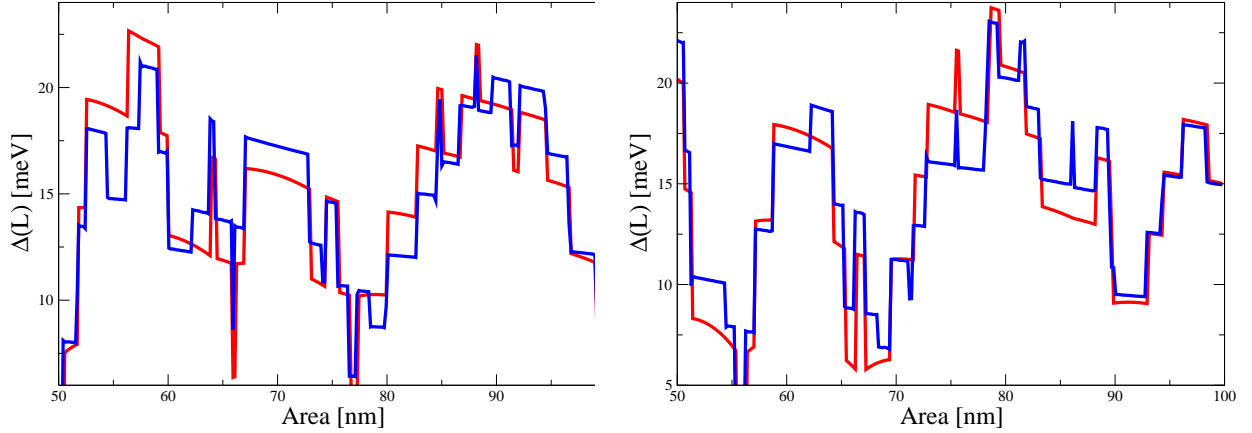

	\centering
	\resizebox{\textwidth}{!}{
		\resizebox{0.49\textwidth}{!}{\includegraphics{gapnumana1_2.eps}}
		\resizebox{0.49\textwidth}{!}{\includegraphics{gapnumana1_4.eps}}}
		\centering
	\caption{Size dependence of the low-temperature superconducting gap: (Blue line) analytical result from Eqs. (\ref{eq:deltaL}), (\ref{eq:bulkgap}), (\ref{eq:f12}) and (\ref{eq:f1}) with $\lambda = 0.22$, $\epsilon_0 = 4$ meV, $k_F =2.06$ nm, $\omega_D=100$ meV and $\epsilon_F = 60$ meV ; (Red line): Numerical evaluation of the gap from the second line of (\ref{eq:Eliashberg_gap_energy}) for the same parameters. Left: nano-island of rectangular shape of aspect ratio $\alpha =1.2$, for all areas.  Right: the same for an aspect ratio $\alpha =1.4$. In both cases we find excellent agreement between numerical and analytical results. We have found a similar agreement for other aspect ratios. The small difference  between the analytical and numerical results is likely due to the fact that we are considering only the leading contribution in $\propto \Delta_0/\omega_D \ll 1$. Higher-order corrections will bring an even better agreement with the numerical results.}
	\label{fig:anagap}
\end{figure}

The integral over energy in the denominator can be solved exactly to give:

\begin{equation}
	\Delta_0^2 \int_{-\epsilon_0}^{\epsilon_0} d\epsilon \frac{1}{(\Delta_0^2 + \epsilon^2)^{3/2}} = 
	\frac{2\epsilon_0}{\sqrt{\Delta_0^2 + \epsilon_0^2}}
	\approx \frac{2\epsilon_0}{\Delta_0}
	\label{eq:denominator}
\end{equation}

\noindent where in the last step we considered the limit $\epsilon_0 \ll \Delta_0$, which was used to derive a closed-form expression for the numerator. Dividing (\ref{eq:numerator}) by (\ref{eq:denominator}) gives $f_{1/2}$:

\begin{equation}
	f_{1/2} = \Big(1 - \frac{3\Delta_0}{2 \omega_D}\Big) \sum_{L_n \neq 0}^{\infty} J_0(k_F L_n) \sinc(L_n / \xi) \label{eq:f12}
\end{equation}

The calculation of the next-to-leading-order term $\propto (k_F L)^{-1}$, highlighted in appendix A, is more convoluted. Here we only state the final result to leading order in $\frac{\Delta_0}{\omega_D}$:

\begin{equation}
	f_1 = -\Big(1 - \frac{3\Delta_0}{2\omega_D}\Big) \Big[ \frac{L_x+L_y}{k_FL^2} + \sum_{i=x, y} \frac{2L_i}{k_F L^2}\sum_{L_n \neq 0}^{\infty}\cos(k_F L_n^{i}) \sinc \Big(\frac{L_n^{i}}{\xi} \Big) \Big] -\frac{3\Delta_0}{2\omega_D} f_{1/2}^2 
 \label{eq:f1}
\end{equation}

\noindent where $L_x = \alpha L$ and $L_y = L/\alpha$ are the sides of the nano-island, with $\alpha > 1$, $L_{n}^{i} = 2nL_i$ is the length of the periodic orbit and $L = \sqrt{L_xL_y}$. Since the STO substrate is a dielectric, Dirichlet boundary conditions were used, hence the minus sign in the Weyl term.

The final expression for the size dependence of the gap (\ref{eq:deltaL}) in the semiclassical limit is obtained from (\ref{eq:bulkgap}), (\ref{eq:f12}) and (\ref{eq:f1}).
At least for FeSe/STO nano-islands \cite{li2015} $k_F \sim 2$ nm and $L \sim 10$ nm, so it is safe to neglect higher orders in the semiclassical expansion. We also stress that these analytical results are only valid in the limits $\epsilon_0 \ll \Delta_0 \ll \omega_D$ and $\lambda \ll 1$. 

We test explicitly the validity of the semiclassical expression (\ref{eq:deltaL}) by comparing it with the numerical calculation of the gap from (\ref{eq:bulk_gap_2}) using the exact spectral density. Results, depicted in Fig. \ref{fig:anagap}, clearly show that the analytical expression is an excellent quantitative approximation for the numerical gap including the complex pattern of oscillations induced by fluctuations of the spectral density. We have focused on the range of parameters describing FeSe/STO, as this is the main goal of the paper. However, with the appropriate modifications, our results are applicable to any weakly coupled superconductor where electron pairing, mediated by phonons or other mechanism, is dominated by forward scattering.

\begin{figure}
	{\epsfig{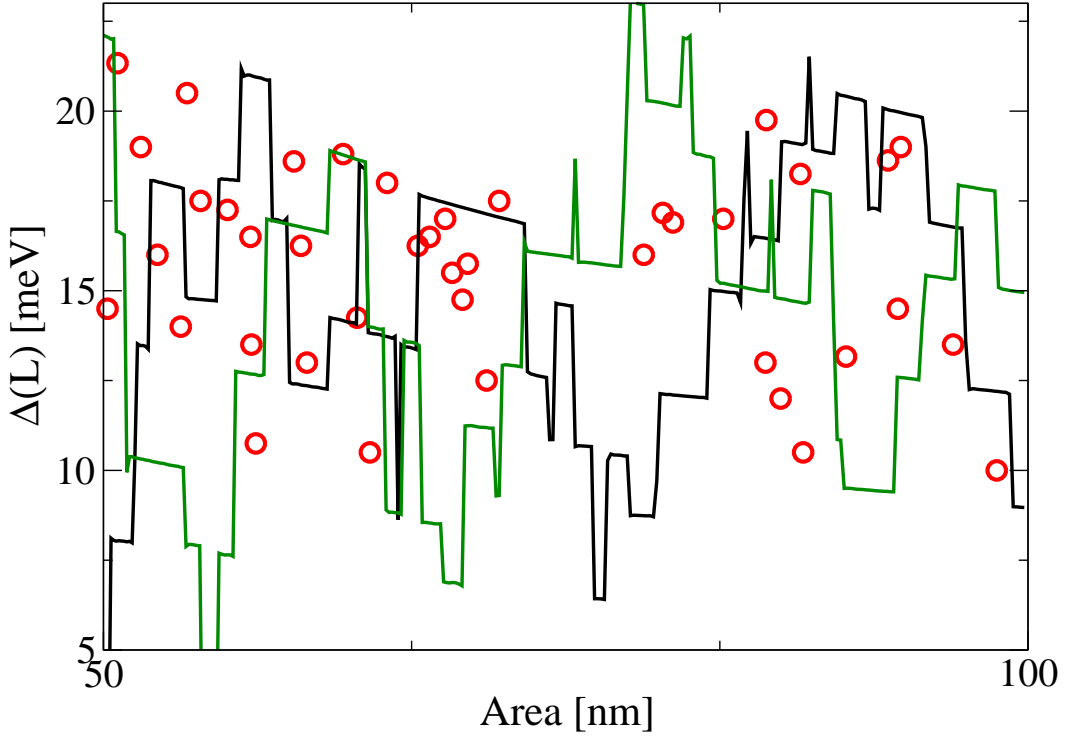}}
	\centering
	\caption{Size dependence of the low-temperature superconducting gap of FeSe nano-islands on a STO substrate: (Black and green lines) analytical result from (\ref{eq:deltaL}), (\ref{eq:bulkgap}), (\ref{eq:f12}) and (\ref{eq:f1}) with $\lambda = 0.22$, $\epsilon_0 = 4$ meV, $k_F =2.06$ nm, $\omega_D=100$ meV and $\epsilon_F = 60$ meV ; (Red circles): Experimental results from Ref. \cite{li2015}. The aspect ratio of the nano-island, which varies from island to island, is not known experimentally but it is expected to be less than $1.5$. We compare the experimental data with the analytical results for two aspect ratios $1.2$ (black) and $1.4$ (green). Similar qualitative agreement is observed for other aspect ratios (not shown). 
	The overall oscillating pattern, including the enhancement of the gap (which can be as large as $40 \%$, for some sizes), is well captured by the analytical expression. For a more quantitative comparison it would be necessary to know experimentally the nano-island aspect ratio.} 
	\label{fig:expgap}
\end{figure}

\section{Comparison with F\lowercase{e}S\lowercase{e}/STO experimental results}
For the sake of clarity, we start by summarizing the range of parameters that are supposed to describe superconductivity in FeSe/STO nano-islands \cite{li2015}. To a good extent the nano-islands are rectangular with area $\sim 50-100$ $\rm {nm}^2$. The aspect ratio varies from island to island and is not known experimentally, but it is expected to belong to the interval $(0,1.5]$. ARPES measurements \cite{lee2014} strongly suggest a Debye energy of $\omega_D \sim 100$ meV. The Fermi energy is of the same order but slightly smaller, $\epsilon_F \sim 60$ meV \cite{gor2015}. Taking into account that 
the effective mass of FeSe electrons is $\rm {m_{\rm eff}} \approx 2.7$ ${\rm m_{\rm e}}$ \cite{zhao2015}, the effective Fermi wavevector $k_F \approx 2$ nm. Assuming forward scattering as the main source of pairing and $\omega_D \sim 100$ meV, an electron-phonon coupling constant of $\lambda \approx 0.2$ is required in order for (\ref{eq:bulkgap}) to reproduce the experimental bulk gap $\Delta_0 \approx 16$ meV.  The phonon spectrum must be strongly peaked around $\omega_D \sim 100$ meV but must still have some finite width, though much smaller than $\omega_D$. Indeed, an exponentially decaying form $|g(\textbf{q})|^2 \propto \exp(-|\textbf{q}| / q_0)$ has been proposed \cite{lee2015}, where $q_0 \sim C/a$ (with $C = \mathcal{O}(1) $ and $a$ is the lattice spacing) is related to the dielectric properties of the FeSe/STO interface.  This matrix element arises from the induced dipole layer generated by the relative displacements of the Ti cations and the oxygen anions in the STO substrate as the phonon modes corresponding to these oscillations are excited. Qualitatively, the cut-off in energy $\epsilon_0$ introduced in the previous section is related to the typical decay on momentum  $\epsilon_0 = \hbar^2 q_0^2 / 2 m^{*} \sim 3$ meV $\ll \Delta_0 \sim 16$ meV.  

Setting $\lambda=0.22, \ k_F=2.06$ nm,$\ \epsilon_0 = 4$ meV, $\omega_D =100$ meV and $\epsilon_F = 60$ meV, we now compare the analytical expression of the gap size dependence (\ref{eq:deltaL}), together with Eqs.(\ref{eq:bulkgap}), (\ref{eq:f12}) and (\ref{eq:f1}), with the experimental results for FeSe nano-islands on a STO substrate \cite{li2015}.
The agreement (see Fig. \ref{fig:expgap}) is reasonably good, especially taking into account that there are no free fitting parameters. Since the aspect ratio varies from nano-island to nano-island, and is not known experimentally (though is expected to be less than $3/2$), we have decided to compare the experimental data with the results for two aspect ratios $1.2$ and $1.4$.
Similar results are obtained for other aspect ratios provided that it is not very close to a square shape. More specifically, as is observed in the figures, the oscillating pattern is sensitive to the aspect ratio but its average deviations from the bulk limit are not. For that reason, and for the sake of clarity, we did not include in Fig.(\ref{fig:expgap}) more analytical results of more aspect ratios. We stress there is no fine tuning of any parameter and the agreement between theory and experiment is in general not very sensitive to small changes of the parameters. 

Our results provide strong evidence that FeSe/STO is mostly a phonon-mediated superconductor where forward scattering is induced by STO phonons with a strongly peaked spectrum around $100$ meV. Although not shown, we have checked that numerical results obtained with more realistic cut-off functions, such as exponential \cite{lee2015}, lead to very similar results by an appropriate rescaling of $\epsilon_0$ still within the allowed range $\epsilon_0 \ll \Delta_0$. We stick to analytical results in order to emphasize the uniqueness of FeSe/STO: a high $T_c$ superconductor that, for the first time,  allows a full analytical quantitative treatment not only of the bulk limit but also of finite size effects. 

In summary, we find a very good agreement between a parameter-free theory and experiments. We stress that, although there is some flexibility, the value of the parameters we use is fixed by experiments or first-principle calculations \cite{lee2014,lee2015}. Additional experiments where the shape of the grains is known with more precision would obviously be helpful to fix other parameters of the model more accurately, including the form of the cut-off function and the value of the electron-phonon coupling constant.

\section{Further enhancement of superconductivity in FeSe/STO}

The experimental results for FeSe/STO nano-islands show an enhancement of the superconducting gap of about $50\%$ for some grain-sizes. Evidently, a single nano-grain $\sim 10$ nm is effectively zero-dimensional so it cannot sustain global long-range order, a distinct feature of a state with zero-resistance. However, a natural question to ask is whether the global critical temperature of a nano-engineered bulk material, composed of an array of these nano-islands connected by Josephson junctions, is enhanced by quantum size effects. This question has been answered affirmatively \cite{mayoh2014,mayoh2015}  in the context of quasi-two dimensional weakly-coupled superconductors. For Al, it was predicted a maximum enhancement of $300\%$ that has recently been confirmed experimentally \cite{pracht2016}. The reason for the enhancement is simply that, although many grains have a low $T_c$, in order for a super-current to exist it is only necessary that a relatively small number (given by the percolation threshold) of grains are still superconducting at the global critical temperature. 

The enhancement that could be achieved in FeSe/STO would likely be much smaller for a number of reasons: 1) the typical length that controls size effects is much smaller than in Al; 2) shell effects are weaker because a rectangular grain has less level degeneracy than spherical Al grains; 3) FeSe/STO is strictly two-dimensional, so quantum and thermal fluctuations, which are detrimental of superconductivity, are stronger. Nevertheless, it is likely that an enhancement of up to $50\%$ \cite{amg} could be observed, provided that it is possible to nano-engineer an array of square (instead of rectangular, as shell effects are stronger in the former) grains of sizes $\sim 6$ nm. Finally, it would be necessary to suppress thermal fluctuations by coupling the interface to a metal or by making the FeSe/STO interface more metallic.

\section{Conclusion}
We have developed a theory of quantum size effects in Eliashberg superconductivity in the limit of weak coupling and peaked phonon spectrum. Our model describes the highly non-monotonic size dependence of the superconducting gap of FeSe/STO nano-islands quantitatively. 
Our results provide further support that FeSe/STO is a weakly-coupled phonon-mediated superconductor with pairing coming from interface phonons with a strongly peaked, but finite, frequency spectrum. Further enhancement of superconductivity is possible by nano-engineering of FeSe/STO superconducting grains.

\acknowledgments
A. M. G. warmly thanks Lili Wang, Canli Song and Zhi Li for providing the experimental data of \cite{li2015} and illumination discussions. A. M. G. acknowledges support
from EPSRC, Grant No. EP/I004637/1.

\appendix

\section{Finite Size Effects for Forward Scattering with a Finite Cut-off}

Starting from equation (\ref{eq:Eliashberg_gap_energy})

\begin{equation}
	1 = \frac{\lambda \omega_D}{2 \epsilon_0 \nu_{TF}(0)} \int_{-\epsilon_0}^{\epsilon_0} d\epsilon \ \nu(\epsilon) \Big( \frac{1}{\sqrt{\epsilon^2 + \Delta^2}} - \frac{3}{2\omega_D}\Big)
\label{eq:Eliashberg_gap_energy_appendix}
\end{equation}

\noindent and expanding the superconducting gap and the density of states, respectively, as $\Delta(L) = \Delta_0 (1 + f_{1/2} + f_1)$ and $\nu(\epsilon) = \nu_{TF}(0) (1 + g_{1/2} + g_1)$, where $f_i$ and $g_i$ are of $\mathcal{O}(k_F L)^{-i}$, including only terms up to $\mathcal{O}(1/k_FL)$:

\begin{equation}
\begin{split}
	1 & = \frac{\lambda \omega_D}{2\epsilon_0} \int_{-\epsilon_0}^{\epsilon_0} d\epsilon (1 + g_{1/2} + g_1) \Bigg[\frac{1}{\sqrt{\epsilon^2 + \Delta_0^2(1 + 2f_{1/2} + 2f_1 + f_{1/2}^2)}} - \frac{3}{2 \omega_D}  \Bigg] = \\
	& =  \frac{\lambda \omega_D}{2\epsilon_0} \int_{-\epsilon_0}^{\epsilon_0} d\epsilon (1 + g_{1/2} + g_1) \Bigg[\frac{1}{\sqrt{\epsilon^2 + \Delta_0^2}} \frac{1}{\sqrt{1 + \frac{\Delta_0^2}{\Delta_0^2 + \epsilon^2} (2f_{1/2} + 2f_1 + f_{1/2}^2})} - \frac{3}{2 \omega_D}  \Bigg] \approx \\
	& \approx \frac{\lambda \omega_D}{2\epsilon_0} \int_{-\epsilon_0}^{\epsilon_0} d\epsilon (1 + g_{1/2} + g_1) \Bigg[\frac{1 - \frac{\Delta_0^2}{\Delta_0^2 + \epsilon^2} (f_{1/2} + f_1 + \frac{f_{1/2}^2}{2}) + \frac{3}{2} (\frac{\Delta_0^2}{\Delta_0^2 + \epsilon_0^2})^2 f_{1/2}^2}{\sqrt{\epsilon^2 + \Delta_0^2}} - \frac{3}{2 \omega_D}  \Bigg] = \\
	& = \frac{\lambda \omega_D}{2\epsilon_0} \int_{-\epsilon_0}^{\epsilon_0} d\epsilon \Bigg[ \Bigg(\frac{1}{\sqrt{\epsilon^2 + \Delta_0^2}} - \frac{3}{2 \omega_D}  \Bigg) \ + \Bigg( g_{1/2} \Big( \frac{1}{\sqrt{\epsilon^2 + \Delta_0^2}} - \frac{3}{2 \omega_D} \Big)  - \frac{\Delta_0^2}{(\epsilon^2 + \Delta_0^2)^{3/2}} f_{1/2} \Bigg) \ + \\
	& \ \ \ + \Bigg( g_{1} \Big( \frac{1}{\sqrt{\epsilon^2 + \Delta_0^2}} - \frac{3}{2 \omega_D} \Big) - \frac{\Delta_0^2}{(\epsilon^2 + \Delta_0^2)^{3/2}} \Big( f_{1} + \frac{f_{1/2}^2}{2} + f_{1/2} g_{1/2} \Big) + \frac{3}{2} \frac{\Delta_0^4}{(\epsilon^2 + \Delta_0^2)^{5/2}} f_{1/2}^2  \Bigg) \Bigg]
\end{split}
\end{equation}

\noindent where after the last equality the terms between the first, second and third pairs of large curly brackets are of $\mathcal{O}(1)$, $\mathcal{O}(1/\sqrt{k_FL})$ and $\mathcal{O}(1/k_F L)$, respectively. In the transition from the second to the third line the binomial expansion $1/\sqrt{1 + x} = 1 - \frac{1}{2} x + \frac{3}{8}x^2 + \mathcal{O}(x^3)$ was carried out, since the corrections $f_i$ and $g_i$ are much smaller than unity.

Equating terms of $\mathcal{O}(1)$ we get:

\begin{equation}
1 = \frac{\lambda \omega_D}{2 \epsilon_0} \int_{-\epsilon_0}^{\epsilon_0} d\epsilon \Big( \frac{1}{\sqrt{\epsilon^2 + \Delta_0^2}} - \frac{3}{2\omega_D}\Big)
\label{eq:bulk_gap_2_appendix}
\end{equation}

which can be easily integrated to lead to obtain an explicit expression for the bulk gap:

\begin{equation}
\Delta_0 = \frac{\epsilon_0}{\sinh \left((1/\lambda+3/2)\frac{\epsilon_0}{\omega_D}\right)} 
\label{eq:bulkgap_appendix}
\end{equation}

Equating terms of $\mathcal{O}(1/\sqrt{k_FL})$ gives:

\begin{equation}
	\large{f_{1/2}} = 
	\frac{\int_{-\epsilon_0}^{\epsilon_0} d\epsilon \ \tilde{g}_{1,2}^{(2)}(\epsilon) \ \Big( \frac{1}{\sqrt{\epsilon^2 + \Delta_0^2}} - \frac{3}{2\omega_D}\Big)}
	{\Delta_0^2 \int_{-\epsilon_0}^{\epsilon_0} d\epsilon \frac{1}{(\Delta_0^2 + \epsilon^2)^{3/2}}}
	\label{eq:correction_2_appendix}
\end{equation}

The numerator can be simplified the following way:

\begin{equation}
	\begin{split}
	& \int_{-\epsilon_0}^{\epsilon_0} d\epsilon \ \tilde{g}_{1,2}^{(2)}(\epsilon) \ \Big( \frac{1}{\sqrt{\epsilon^2 + \Delta_0^2}} - \frac{3}{2\omega_D}\Big) = \\
	& = \int_{-\epsilon_0}^{\epsilon_0} d\epsilon \ \sum_{L_n \neq 0}^{\infty} J_0(k(\epsilon) L_n^{1,2}) \ \Big( \frac{1}{\sqrt{\epsilon^2 + \Delta_0^2}} - \frac{3}{2\omega_D}\Big) \approx \\
	& \approx \int_{-\epsilon_0}^{\epsilon_0} d\epsilon \ \sum_{L_n \neq 0}^{\infty} \sqrt{\frac{2}{\pi k_F L_n^{1,2}}} \cos \Big( k_F(1 + \frac{\epsilon}{2 \epsilon_F}) L_n^{1,2} - \frac{\pi}{4}\Big) \ \Big( \frac{1}{\sqrt{\epsilon^2 + \Delta_0^2}} - \frac{3}{2\omega_D}\Big) \approx \\
	& \approx \sum_{L_n \neq 0}^{\infty} J_0(k_F L_n^{1,2}) \int_{-\epsilon_0}^{\epsilon_0} d\epsilon \ \cos \Big(\frac{k_F\epsilon}{2 \epsilon_F} L_n^{1,2} \Big) \ \Big( \frac{1}{\sqrt{\epsilon^2 + \Delta_0^2}} - \frac{3}{2\omega_D}\Big) \approx \\
	& \approx \sum_{L_n \neq 0}^{\infty} J_0(k_F L_n^{1,2})  \Big( \frac{1}{\Delta_0} - \frac{3}{2\omega_D}\Big) \int_{-\epsilon_0}^{\epsilon_0} d\epsilon \ \cos \Big(\frac{k_F\epsilon}{2 \epsilon_F}  L_n^{1,2} \Big) = \\
	& = \sum_{L_n \neq 0}^{\infty} J_0(k_F L_n^{1,2})  \Big( \frac{1}{\Delta_0} - \frac{3}{2\omega_D}\Big) 2 \frac{2 \epsilon_F}{k_F  L_n^{1,2}} \ \sin \Big(\frac{k_F\epsilon_0}{2 \epsilon_F} L_n^{1,2} \Big) \equiv \\
	& \equiv  \sum_{L_n \neq 0}^{\infty} J_0(k_F L_n^{1,2})   2\epsilon_0 \Big( \frac{1}{\Delta_0} - \frac{3}{2\omega_D}\Big) \ \sinc \Big(\frac{L_n^{1,2}}{\xi} \Big)
	\end{split}
\label{eq:numerator_appendix}
\end{equation}

\noindent where $\sinc(x) \equiv \sin(x) / x$ and $\xi \equiv \frac{2\epsilon_F}{k_F \epsilon_0}$ is the relevant coherence length. In the transition from the second to the third line, the asymptotic limit $J_0(x) = \sqrt{\frac{2}{\pi x}} \cos(x - \frac{\pi}{4})$ was used and $k(\epsilon)$ was expanded about the Fermi wavevector $k_F$. In the following line, the double-angle formula $\cos(a + b) = \cos(a) \cos(b) - \sin(a) \sin(b)$ was used and the term involving the sines was neglected since $\sin(\frac{k_F\epsilon}{2 \epsilon_F} L_n^{1,2}) \ll 1$. Given that $\epsilon_0 \ll \Delta_0 \ll \omega_D$, the term between curly brackets in the integrand was assumed constant. However, since the periodic orbits $L_n^{1,2}$ can be arbitrarily large, the change of the phase of the cosine over the range of integration cannot be neglected.

The denominator can also be evaluated explicitly, 

\begin{equation}
	\Delta_0^2 \int_{-\epsilon_0}^{\epsilon_0} d\epsilon \frac{1}{(\Delta_0^2 + \epsilon^2)^{3/2}} = \frac{2 \epsilon_0}{\sqrt{\epsilon_0^2 + \Delta_0^2}} \approx \frac{2\epsilon_0}{\Delta_0}
\label{eq:denominator_appendix}
\end{equation}

\noindent where in the last step we considered the limit $\epsilon_0 \ll \Delta_0$, which was used to derive a closed-form expression for the numerator. Dividing (\ref{eq:numerator_appendix}) by (\ref{eq:denominator_appendix}) gives the leading-order correction:

\begin{equation}
	f_{1/2} = \Big(1 - \frac{3\Delta_0}{2 \omega_D}\Big) \sum_{L_n \neq 0}^{\infty} J_0(k_F L_n) \sinc(L_n / \xi) \label{eq:f12_appendix}
\end{equation}

Equating terms of $\mathcal{O}(1/k_FL)$:

\begin{equation}
	\begin{split}
	& \Big[ \int_{-\epsilon_0}^{\epsilon_0} d\epsilon \frac{\Delta_0^2}{(\epsilon^2 + \Delta_0^2)^{3/2}} \Big] f_{1} = \\
	& \int_{-\epsilon_0}^{\epsilon_0} d\epsilon \Bigg( g_{1} \Big( \frac{1}{\sqrt{\epsilon^2 + \Delta_0^2}} - \frac{3}{2 \omega_D} \Big) - \frac{\Delta_0^2}{(\epsilon^2 + \Delta_0^2)^{3/2}} \Big(\frac{f_{1/2}^2}{2} + f_{1/2} g_{1/2} \Big) + \frac{3}{2} \frac{\Delta_0^4}{(\epsilon^2 + \Delta_0^2)^{5/2}} f_{1/2}^2  \Bigg)
	\end{split}
\end{equation}

\noindent where $g_1 = \overline{g}(0) - \frac{1}{2} \sum_{i}\tilde{g}_{i}^{(1)}(\epsilon) = - \frac{L_x+L_y}{k_FL^2} - \sum_{i=x, y} \frac{2L_i}{k_F L^2}\sum_{L_n \neq 0}^{\infty}\cos(k(\epsilon)L_n^{i})$ is the $\mathcal{O}(1/k_F L)$ correction to the density of states and, as before, $g_{1/2} = \sum_{L_n \neq 0}^{\infty} J_0(k(\epsilon) L_n^{1,2})$. The term in the Left-Hand Side (LHS) and the two terms involving $f_{1/2}^2$ in the Right-Hand Side (RHS) can be simplified by solving the integrals assuming $\epsilon_0 \ll \Delta_0$:

\begin{equation}
	\frac{2\epsilon_0}{\Delta_0} f_{1} = \int_{-\epsilon_0}^{\epsilon_0} d\epsilon \Bigg( g_{1} \Big( \frac{1}{\sqrt{\epsilon^2 + \Delta_0^2}} - \frac{3}{2 \omega_D} \Big) - \frac{\Delta_0^2}{(\epsilon^2 + \Delta_0^2)^{3/2}} f_{1/2} g_{1/2} \Bigg) + \frac{2\epsilon_0}{\Delta_0} f_{1/2}^2 \Big( \frac{3}{2} - \frac{1}{2} \Big)
	\label{eq:term1}
\end{equation}

The first term on the RHS can be written in closed-form as:

\begin{equation}
	\begin{split}
	& - \int_{-\epsilon_0}^{\epsilon_0} d\epsilon \ \Big(\frac{L_x+L_y}{k_FL^2} + \sum_{i=x, y} \frac{2L_i}{k_F L^2}\sum_{L_n \neq 0}^{\infty}\cos(k(\epsilon)L_n^{i}) \Big) \Big( \frac{1}{\sqrt{\epsilon^2 + \Delta_0^2}} - \frac{3}{2 \omega_D} \Big) \approx \\
	& \approx - 2 \epsilon_0 \Big(\frac{1}{\Delta_0} - \frac{3}{2\omega_D} \Big) \Big[\frac{L_x+L_y}{k_FL^2} + \sum_{i=x, y} \frac{2L_i}{k_F L^2}\sum_{L_n \neq 0}^{\infty}\cos(k_F L_n^{i}) \sinc(L_n^{i} / \xi) \Big]
	\end{split}
	\label{eq:term2}
\end{equation}

\noindent where in the transition from the first to the second line the approximations described in (\ref{eq:numerator_appendix}) were used. The term $\sim f_{1/2}g_{1/2}$ can also be simplified:

\begin{equation}
	\begin{split}
	& - \int_{-\epsilon_0}^{\epsilon_0} d\epsilon \frac{\Delta_0^2}{(\epsilon^2 + \Delta_0^2)^{3/2}} f_{1/2} g_{1/2} = \\
	& = - f_{1/2} \int_{-\epsilon_0}^{\epsilon_0} d\epsilon \frac{\Delta_0^2}{(\epsilon^2 + \Delta_0^2)^{3/2}} \sum_{L_n \neq 0}^{\infty} J_0(k(\epsilon) L_n^{1,2}) \approx \\
	& \approx - f_{1/2} \sum_{L_n \neq 0}^{\infty} J_0(k_F L_n^{1,2}) \int_{-\epsilon_0}^{\epsilon_0} d\epsilon \ \frac{\Delta_0^2}{(\epsilon^2 + \Delta_0^2)^{3/2}} \cos \Big(\frac{k_F\epsilon}{2 \epsilon_F} L_n^{1,2} \Big) \approx \\
	& \approx - f_{1/2} \frac{1}{\Delta_0} \sum_{L_n \neq 0}^{\infty} J_0(k_F L_n^{1,2}) \int_{-\epsilon_0}^{\epsilon_0} d\epsilon \ \cos \Big(\frac{k_F\epsilon}{2 \epsilon_F} L_n^{1,2} \Big) = \\
	& = - f_{1/2} \frac{2\epsilon_0}{\Delta_0} \sum_{L_n \neq 0}^{\infty} J_0(k_F L_n^{1,2}) \sinc(L_n^{1,2} / \xi) = \\
	& = - \frac{\frac{2\epsilon_0}{\Delta_0}}{\Big(1 - \frac{3\Delta_0}{2 \omega_D}\Big)} f_{1/2}^2
	\end{split}
	\label{eq:term3}
\end{equation}

\noindent where again all steps were previously described in (\ref{eq:numerator_appendix}). Combining Eqs. (\ref{eq:term1}), (\ref{eq:term2}) and (\ref{eq:term3}) gives the next-to-leading-order correction:

\begin{equation}
	f_1 = -\Big(1 - \frac{3\Delta_0}{2\omega_D}\Big) \Big[ \frac{L_x+L_y}{k_FL^2} + \sum_{i=x, y} \frac{2L_i}{k_F L^2}\sum_{L_n \neq 0}^{\infty}\cos(k_F L_n^{i}) \sinc \Big(\frac{L_n^{i}}{\xi}\Big) \Big] + f_{1/2}^2 \Big( 1 - \frac{1}{1 - \frac{3\Delta_0}{2\omega_D}} \Big)
\end{equation}

\bibliography{library}
\end{document}